\documentclass[a4paper]{jpconf}
\usepackage{graphicx}
\begin{document}
\title{Recent Results from MINER$\nu$A}

\author{Laura Fields, for the MINER$\nu$A collaboration}

\address{Fermi National Accelerator Laboratory, PO Box 500, Batavia, IL 60510-5011, USA}

\ead{ljf26@fnal.gov}
\medskip
\address{FERMILAB-CONF-16-393-CD-ND}

\begin{abstract}
The MINER$\nu$A collaboration is currently engaged in a broad program
of neutrino-nucleus interaction measurements.  Several
recent measurements of interest to the accelerator-based oscillation
community are presented.  These include measurements of quasi-elastic
scattering, diffractive pion production, kaon
production and comparisons of interaction cross sections across nuclei.  A new
measurement of the NuMI neutrino beam flux that incorporates both
external hadro-production data and MINER$\nu$A detector data is also presented.  
\end{abstract}

\section{Introduction}
\begin{figure} [h]
\begin{center}
\includegraphics[width=0.4\textwidth]{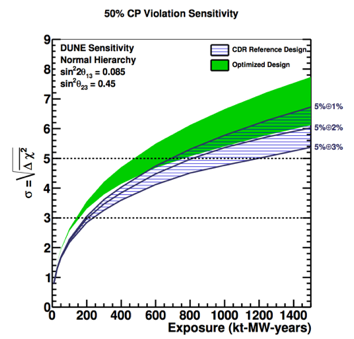}
\end{center}
\caption{\label{fig:dune}Estimated sensitivity of the DUNE experiment to 50\% of the possible
  values $\delta_{CP}$, versus exposure for two different beam options
 and assuming systematic uncertainties on the $\nu_e$ signal relative
 to other species of 1\% (top), 2\% (middle) and 3\% (bottom)~\cite{dune2}.}
\end{figure}

Reduction of systematic uncertainties is becoming
increasingly critical to accelerator-based oscillation measurements.
Figure~\ref{fig:dune} shows the CP sensitivity of the DUNE experiment as a
function of exposure for several systematics goals, and illustrates
that the difference between a 1\% and 3\% systematic uncertainty on
the $\nu_e$ signal relative to other neutrino species is equivalent to nearly doubling the
exposure required to reach 5$\sigma$ coverage of 50\% of $\delta_{CP}$
phase space~\cite{dune2}.  Reaching such low systematics goals will require control
of all systematics, including those from neutrino interaction cross
sections, flux and detector response.

Reduction of systematics arising from neutrino cross-section
parameters is complicated by the fact that a single cross section measurement simultaneously measures
dozens of parameters, describing both the underlying nucleon
cross section and initial and final state nuclear effects.
Untangling these many parameters can only be achieved through many
independent measurements, of many different interaction channels, at
various energies, on different nuclei and with varying reconstruction
techniques. 

The MINER$\nu$A experiment~\cite{minerva} was designed to make such measurements.
Composed of plastic scintillator strips interspersed with other
materials, MINER$\nu$A sits in the NuMI beam at Fermilab and has
accumulated large datasets in both the low energy and medium energy
configurations of the NuMI beam.  As of this writing, the MINER$\nu$A
collaboration has produced fifteen neutrino cross-section
publications, and is hard at work on many more.  A selection of these
results are described in detail here.

\section{The NuMI Flux}
\begin{figure} [h]
\begin{center}
\includegraphics[width=0.4\textwidth]{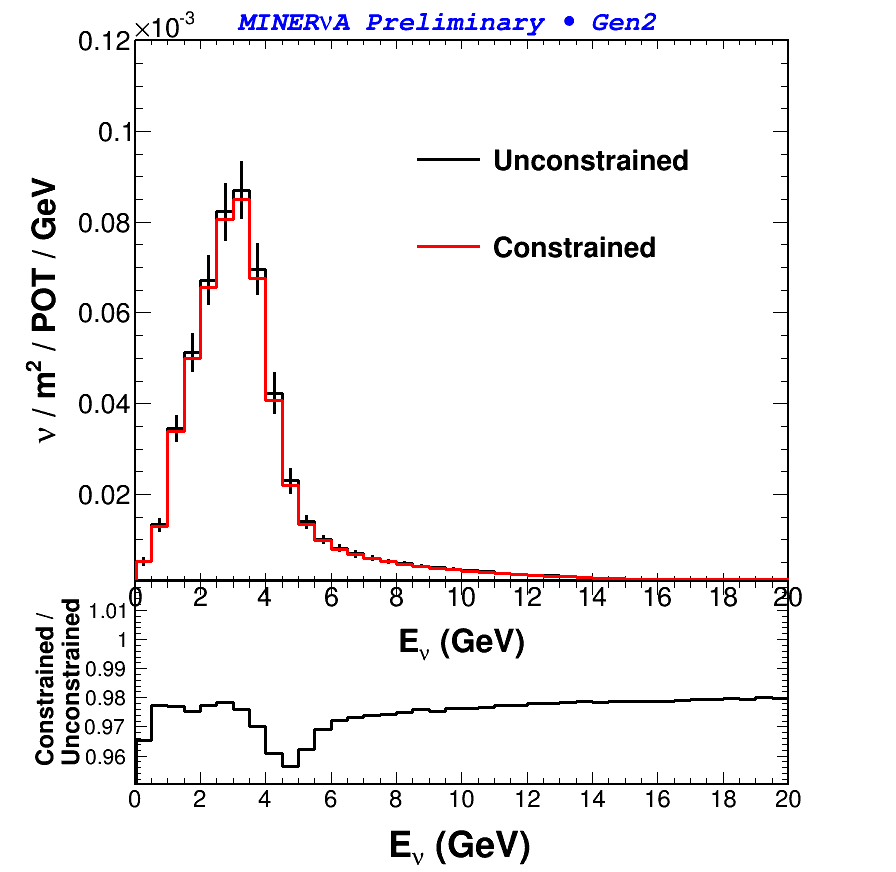}
\includegraphics[width=0.4\textwidth]{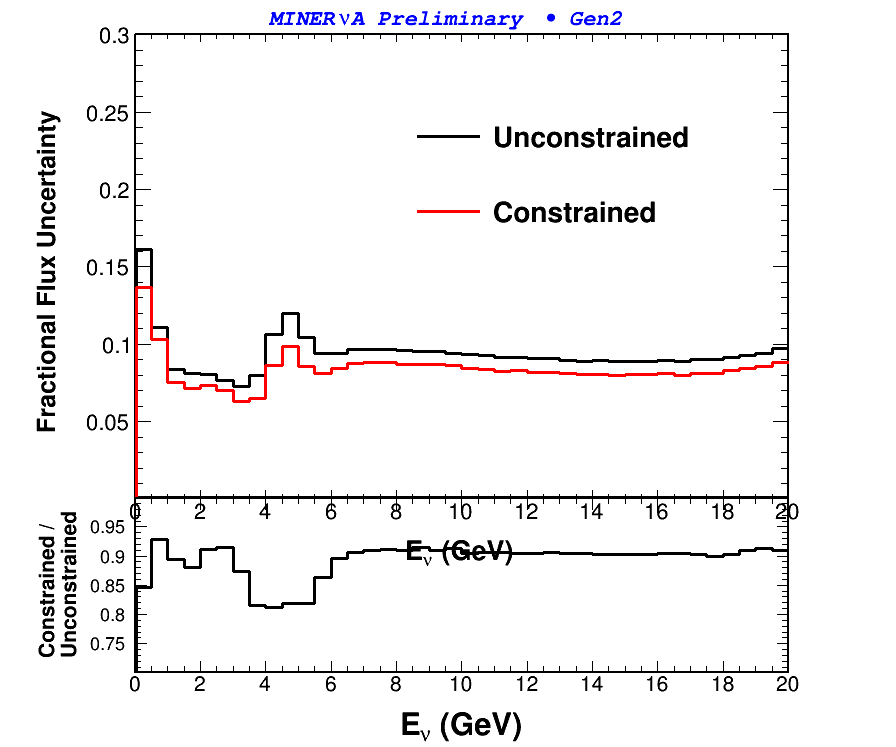}
\end{center}
\caption{\label{fig:flux}Predicted NuMI low energy, muon neutrino
  flux (left) and flux uncertainties (right) at MINER$\nu$A before and
  after constaint with neutrino-electron scattering data.~\cite{leo}.}
\end{figure}

A key ingredient in all of MINER$\nu$A's cross section measurements is a
prediction of the number, flavor and energy spectrum of neutrinos in
the NuMI beamline.  An estimate of the neutrino fluxes at MINER$\nu$A
using external hadro-production data was recently made~\cite{leo}, and achieved
flux uncertainties of approximately 8\% in the focusing peak.
MINER$\nu$A has further demonstrated the use of neutrino-electron
scattering as a standard candle for flux measurement~\cite{jaewon}.  Because
neutrino-electron scattering has an extremely small cross section,
only 135 events (after background subtraction and acceptance
correction) were identified in MINER$\nu$A's complete low energy
neutrino run.  Combining the measured electron energy spectrum of
these events with
the {\it a priori} flux uncertainty based on external measurements
reduces the total flux uncertainty in the focusing peak form 8\% to
7\%, as illustrated in Figure~\ref{fig:flux} 

\section{Quasi-elastic Scattering}
Among MINER$\nu$A's most important measurements are those of
quasi-elastic scattering.  Dominating the charged-current cross
section near 1 GeV and having a relatively simple final state, these
processes are considered a golden channel for oscillation
experiments.  In recent years, it has become clear that measurements
of this process are complicated by ``quasielastic-like'' final states
that can arise both from pion production processes wherein a pion is
absorbed before exiting the nucleus, and by neutrino interactions with
multi-nucleon bound states.

\begin{figure}
\begin{center}
\includegraphics[width=0.75\textwidth]{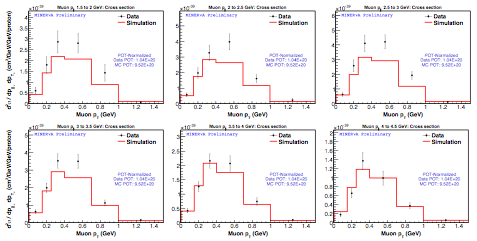}
\end{center}
\caption{\label{fig:cheryl} Double differential muon neutrino charged current
  quasi-elastic scattering cross sections versus muon longitudinal and
transverse momentum.}
\end{figure}

MINER$\nu$A's program of quasi-elastic measurements demonstrates the
experiment's ability to study a process from multiple perspectives.
MINER$\nu$A's first measurements were of single differential muon
neutrino and antineutrino cross section versus $Q^2$ (the momentum
transferred from the initial state neutrino to the final state
nucleon).  Those studies are currently being expanded to measurements
double-different cross sections with respect to muon transverse and
longitudinal momentum.  A preliminary version of the antineutrino
measurement is shown in Figure~\ref{fig:cheryl}, with MINER$\nu$A data
indicating significant additional strength at moderate transverse
momentum when compared to the GENIE~\cite{genie} event generator.

\begin{figure}[h]
\begin{center}
\includegraphics[width=0.6\textwidth]{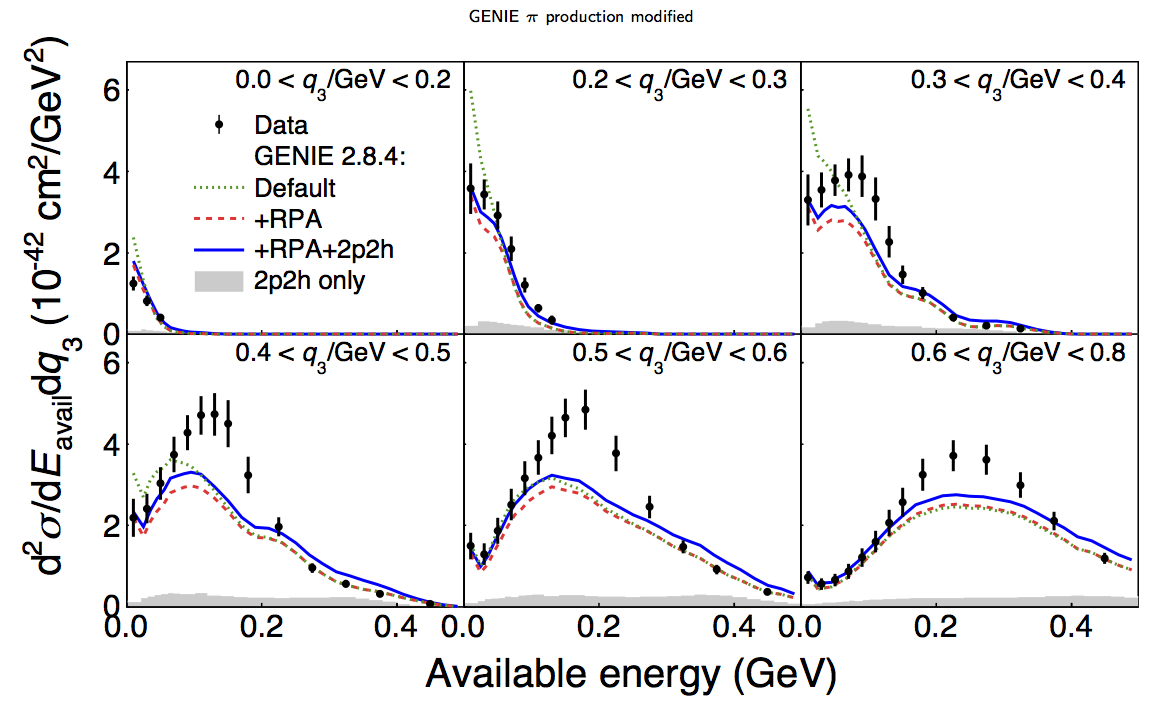}
\end{center}
\caption{\label{lowrecoil}Double differential cross sections of charged 
  current inclusive scattering as a function of $q^3$ and available 
  energy~\cite{phil}.  Available energy is the sum of proton and charged pion kinetic
energy and neutral pion, electron, and photon total energy.}
\end{figure}

Quasi-elastic and quasi-elastic final states have also been studied
through a measurement of inclusive
charged-current scattering with low recoil energy.  Results of this
analysis are reported in~\cite{phil} and shown in
figure~\ref{lowrecoil}.  Significant discrepancies with
GENIE models with and without two-particle, two-hole additions are
found in regions of moderate energy.  The excess in this inclusive
neutrino analysis appears in a similar kinematic region to the
exclusive antineutrino quasi-elastic analysis discussed above.  

\begin{figure}[h]
\begin{center}
\includegraphics[width=0.5\textwidth]{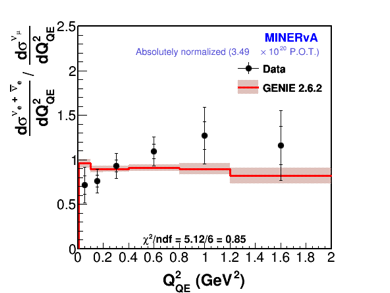}
\end{center}
\caption{Ratio of electron neutrino and muon neutrino
  charged-current quasi-elastic cross sections as a function of
  $Q^2$~\cite{jeremy1}.}
\label{fig:electronCCQE}
\end{figure}

Although muon neutrinos and antineutrinos make up the majority of the
NuMI beam, there is a small ($\sim 1$\%) component of electron
neutrinos.  MINER$\nu$A has used this portion of the flux to study
electron neutrino quasi-elastic scattering.  Of particular interest is
the ratio of electron neutrino to muon neutrino scattering.
Uncertainties on electron neutrino scattering that are uncorrelated
with the corresponding muon neutrino processes (which can arise from
unknown nuclear effects coupling to known differences in the cross
sections due to differences between the electron and muon mass)
are amongst the most critical to constrain for future oscillation
measurements.  The ratio of the electron neutrino CCQE cross section
versus $Q^2$ is shown in Figure~\ref{fig:electronCCQE}.  Within
uncertainties, the GENIE neutrino event generator models
differences between these two processes well.

\section{Diffractive Neutral Current Pion Production}

\begin{figure}[h]
\begin{center}
\includegraphics[width=0.5\textwidth]{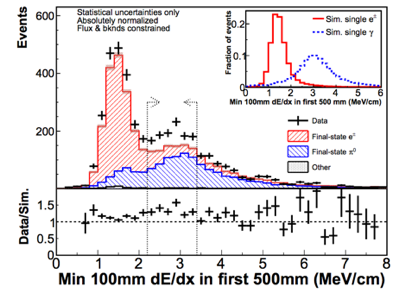}
\end{center}
\caption{Energy deposition per unit length at the
  upstream end of candidate electron showers in the electron neutrino
  CCQE analysis.  The region between the arrows defines the ``excess
  region'' containing unmodeled events consistent with neutral current
  diffractive scattering.  The region to the left of the arrows is the
  signal region in the CCQE analysis \cite{jeremy2}.}
\label{fig:jeremy}
\end{figure}

The dE/dx distribution at the upstream end of the electron candidate
in the electron neutrino CCQE analysis described in the previous
section is shown in Figure~\ref{fig:jeremy}.  The signal region peaked
near 1.5 MeV/cm corresponds to the energy deposition
expected for a minimum ionizing particle.  In the sideband between 2.2
and 3.5 GeV, an excess beyond the Monte Carlo prediction is observed
consistent with two overlapping minimum ionizing particles.  This
excess of events has been studied and found to have a signature that
is similar to coherent pion production, in that the electromagnetic
showers are very forward, and there is little other energy present in
the event.  However, when considering energy in a cone extending
upstream from the shower vertex, these excess events contain
significantly more energy than that simulated coherent events.  This
energy is consistent with 
diffractive pion production from Hydrogen, a process that is not
simulated in GENIE.

\section{Kaon Production}
Neutrino induced pion production is of interest because it is a
background to the SUSY-preferred proton decay channel $p\rightarrow
K^{+}\nu$.  This is a particularly problematic background for water
Cherenkov detectors, where the kaon is typically below the Cherenkov threshold.
Kaon production by neutrinos is also a sensitive test of kaon FSI
models, which will be necessary for discovering proton decay in any
detector, including liquid Argon detectors such as DUNE.  

\begin{figure}[h]
\begin{center}
\includegraphics[width=0.4\textwidth]{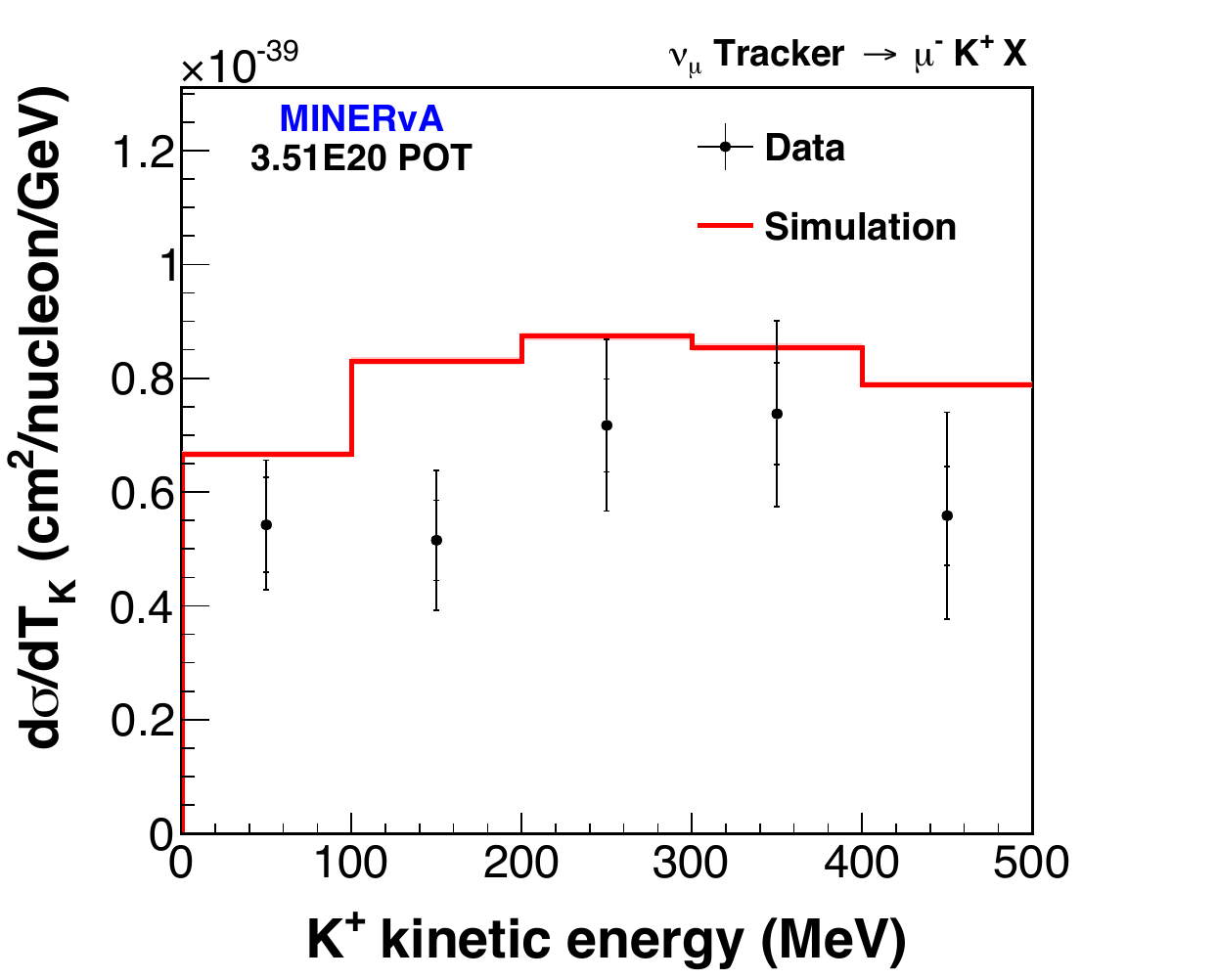}
\includegraphics[width=0.4\textwidth]{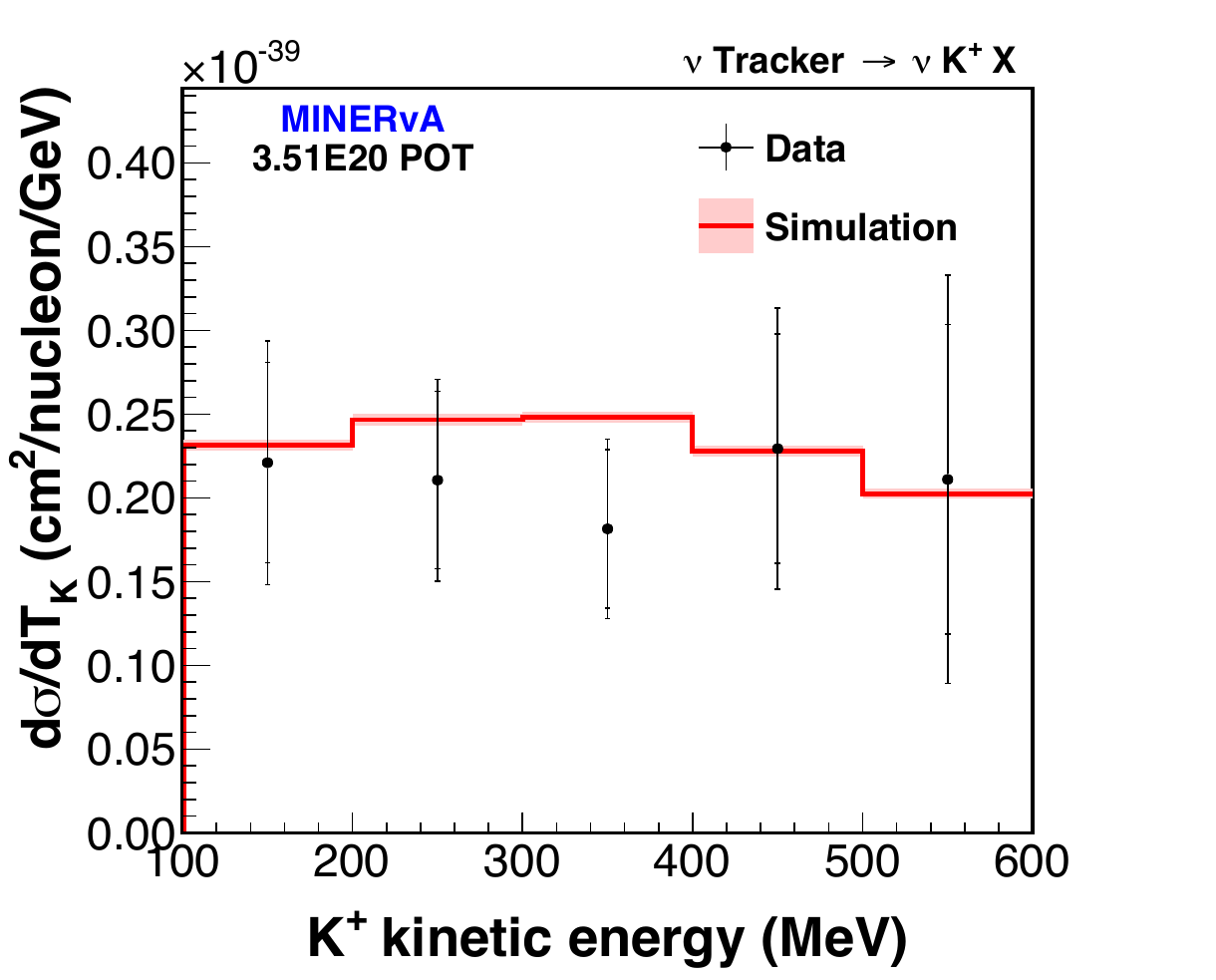}
\end{center}
\caption{\label{chris}Differential cross sections of charged 
  current (left) and neutral current (right) kaon production as a 
  function of kaon kinetic energy~\cite{chris1}.}
\label{fig:chris}
\end{figure}

MINER$\nu$A's nanosecond-level timing resolution enables
identification of kaon production via observation of the time delay
between a kaon track and its decay products.  This technique has been
used to measure cross sections for both neutral current and charged
current kaon production, shown in Figure~\ref{fig:chris}.  In both
cases, good agreement is found with the GENIE event generator, in
spite of known deficiencies in that model, including the
omission of single pion production.  These data indicate that, as
those deficiencies are addressed, the overall rate of kaon production
should not be substantially changed~\cite{chris1}.

\section{Comparison of Cross Sections across Different Nuclei}
\begin{figure}[h]
\begin{center}
\includegraphics[width=0.5\textwidth]{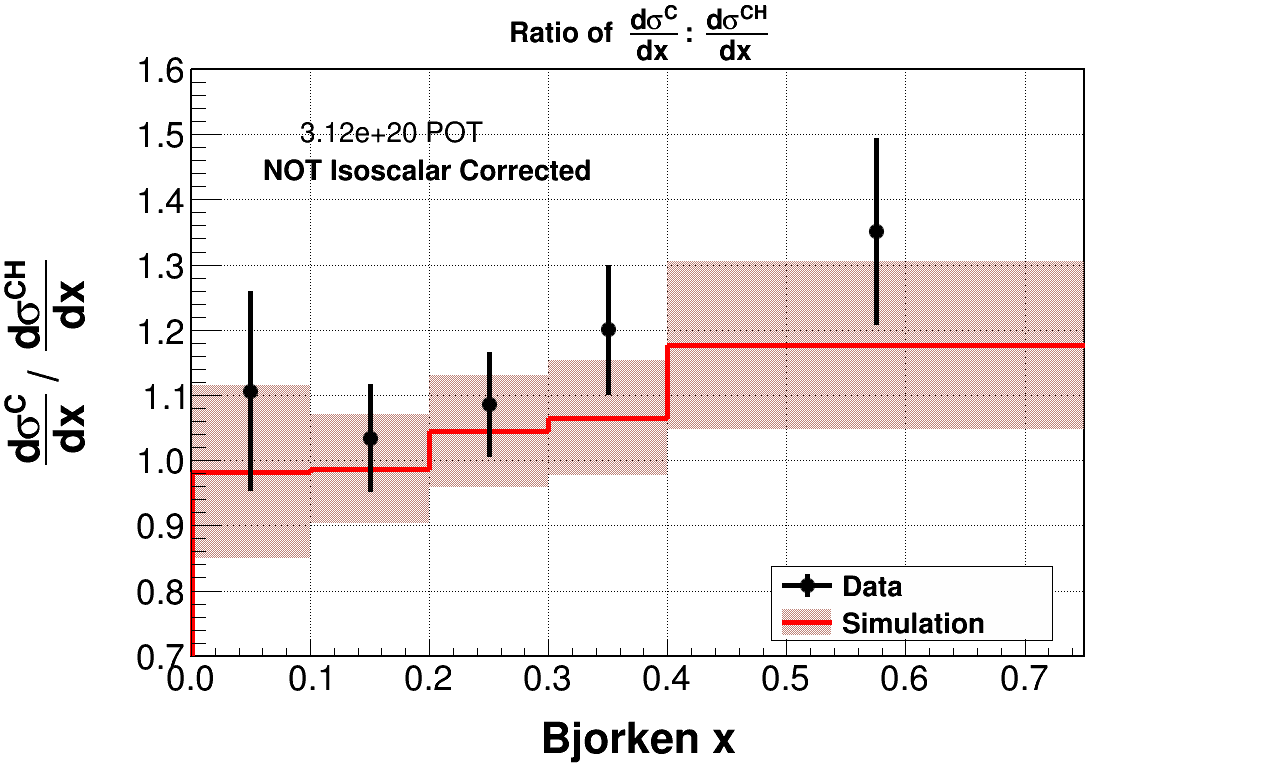}
\includegraphics[width=0.5\textwidth]{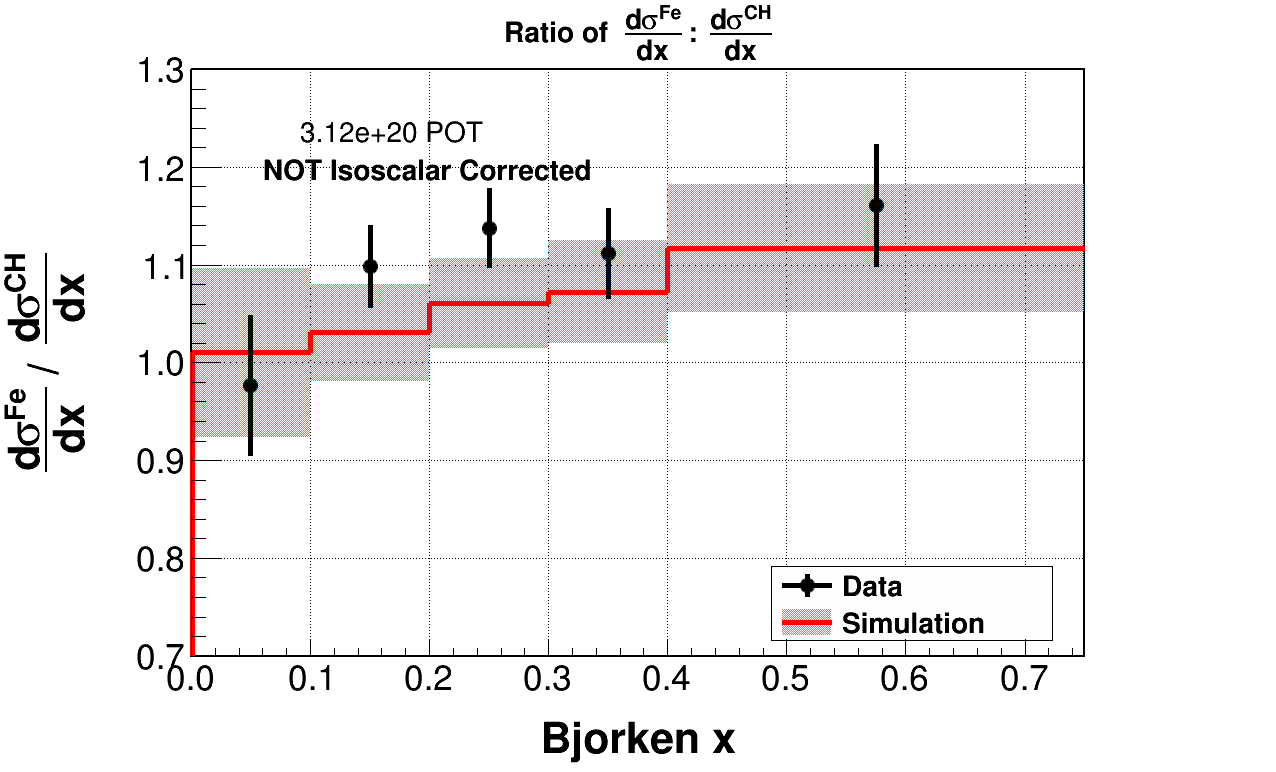}
\includegraphics[width=0.5\textwidth]{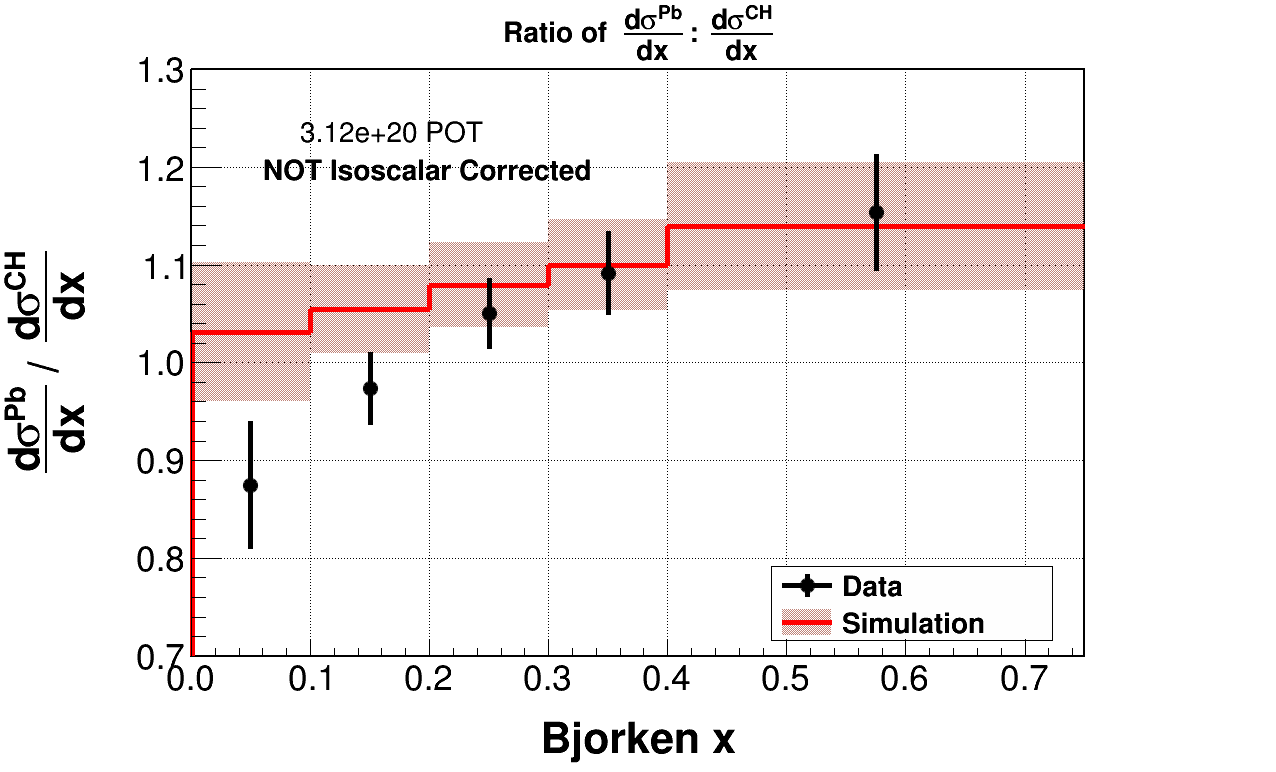}
\end{center}
\caption{Ratios to scintillator of carbon (left), iron (middle) and lead (right) differential cross sections of charged 
  current deep inelastic scattering as a 
  function of function of Bjorken $x$.~\cite{joel}.}
\label{fig:ratios}
\end{figure}
An important feature of the MINER$\nu$A detector is the ability to
measure cross sections as a function of different nuclei.  Recently,
MINER$\nu$A has studied deep inelastic scattering as a function of
Bjorken $x$ on Carbon, Iron, Lead and scintillator.  Ratios of these
cross sections are shown in Figure~\ref{fig:ratios}.  A small unmodeled deficit
in heavy nuclei at low $x$, the so called ``shadowing region'' is
observed here and mirrors an earlier measurement of charged current
inclusive cross section ratios~\cite{brian}.

\section{Prospects for the Future}
\begin{figure}
\begin{center}
\includegraphics[width=0.65\textwidth]{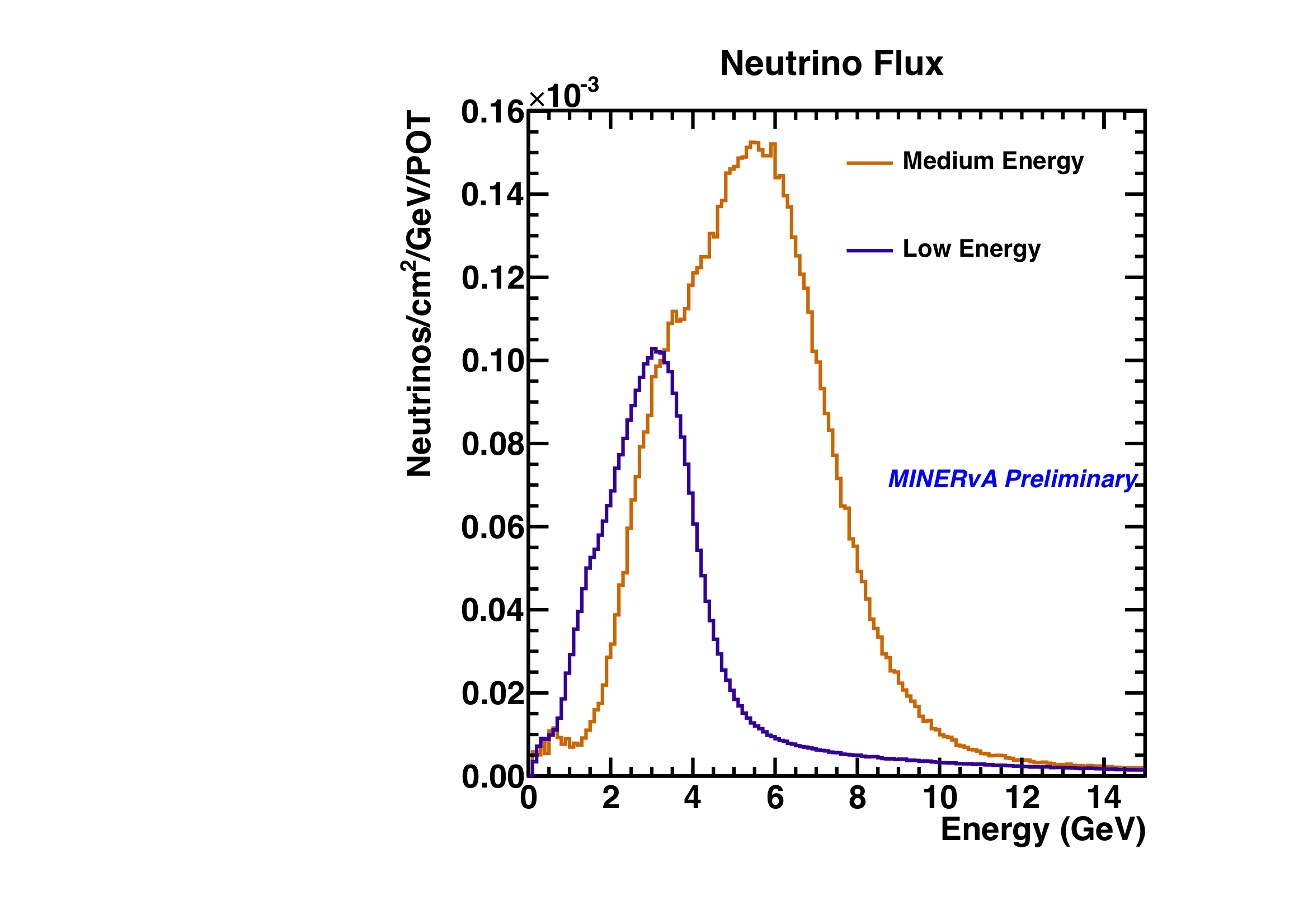}
\end{center}
\caption{Comparison of simulated neutrino-mode muon neutrino fluxes at 
  MINER$\nu$A in the low and medium energy NuMI beam tunes.}   
\label{fig:flux2}
\end{figure}

With fifteen publications so far, MINER$\nu$A is nearing the
completion of analysis of its low energy datasets.  Attention is now
turning to a large dataset currently being accumulated in the
medium energy tune of the NuMI beamline.  This beam provides a higher
flux beam with spectrum shown in Figure~\ref{fig:flux2}.  A factor of
three more protons-on-target than were used in the analyses discussed
above have already been accumulated, and an antineutrino dataset is
also expected.  These will provide a much stronger flux constraint
from neutrino-electron scattering and will facilitate high statistics
comparisons of exclusive channels across nuclei.  

\section*{Acknowledgements}
This work was supported by the Fermi National Accelerator Laboratory under US Department
of Energy contract No. DE-AC02-07CH11359 which included the MINERvA construction
project. Construction support also was granted by the United States National
Science Foundation under Award PHY-0619727 and by the University of Rochester. Support
for participating scientists was provided by NSF and DOE (USA) by CAPES and CNPq
(Brazil), by CoNaCyT (Mexico), by CONICYT (Chile), by CONCYTEC, DGI-PUCP and
IDI/IGI-UNI (Peru), by Latin American Center for Physics (CLAF). We thank the MINOS
Collaboration for use of its near detector data. Finally, we thank Fermilab for support of
the beamline and the detector, and in particular the Scientific Computing Division and the
Particle Physics Division for support of data processing.

\end{document}